# Third parameter classification of transients and novae ejecta as ballistically ejected globules

**Robert Williams**

Space Telescope Science Institute
Baltimore, MD  21218   USA

wms@stsci.edu

**Abstract**. A third parameter, in addition to luminosity and rate of brightness decline, derived from the spectra of transients is suggested as a means of more accurately classifying objects in outburst. Principal component analysis of the spectra of transients is suggested as the best way to determine the third parameter. A model is suggested for novae ejecta that is based on ballistic ejection of an ensemble of globules. The distribution in sizes of the clouds should determine basic characteristics of the ejecta, including the short term brightness fluctuations of novae in decline and the location of dust formation.

## 1. Introduction

Although the basics of the nova outburst have been understood for some time there is still uncertainty about some aspects of the phenomenon. Outbursts are the result of steady accretion but there may be other triggers. What causes many novae to vary 1-2 magnitudes in brightness over a few weeks time during decline? What are the key parameters that determine how and when some novae develop substantial dust? In recent years new luminous transients have been discovered that do not fit into either the normal nova or supernova paradigm. It is important to determine what observations may be most useful in understanding the physical mechanism driving these outbursts.

## 2. Classification of Transients

The standard plot used to describe transients is one of luminosity vs. brightness decline time. The significance of this plot derives from Arp's classic study of the novae in M31 [1], which he undertook immediately after completing his doctoral thesis at Caltech with the support of W. Baade, A. Sandage, and E. Hubble shortly before the latter's death. Observing 290 clear nights on Mt. Wilson over a 1.5 yr interval in a Herculean effort Arp found that rapidly declining novae were more luminous than slow decliners. The resulting Maximum Magnitude vs. Rate of Decline (MMRD) relation, calibrated independently from the Cepheid, RR Lyrae, H II region distances for M31, was useful to Sandage as a more refined primary extragalactic distance indicator compared to the less accurate average absolute magnitude of the few novae determined by Curtis [2] in his famous debate with Shapley over the size of the Milky Way and distance to M31.

The MMRD relation has an intrinsic dispersion that has steadily increased over time with the larger sample of novae that have been observed within nearby galaxies since Arp's original survey. Until the past two decades a plot of visible luminosity vs. time of decline for transients was dominated primarily by novae and SNe that occupied a relatively small region of space in the plot. However, as seen in figure 1 whose data come primarily from the Palomar Transient Factory survey [3,4], the MMRD relation is not very reliable, with many variant outlier novae. And transients of all types occupy virtually every part of the luminosity-decay time diagram, as can be seen in figure 2.

Historically, the position of a transient on a plot of maximum magnitude vs. light decay time has been a standard way of expressing the nature or type of the transient. However, it is clear from figure 2 that there is so much overlap of different transient types in the diagram that little information on the nature of a transient results from its position in this diagram. Since many transients have a broad distribution of decay times following outburst the decline time is not much of a defining characteristic of a transient. Rather, some other parameter would be much more useful in distinguishing between different transient phenomena.

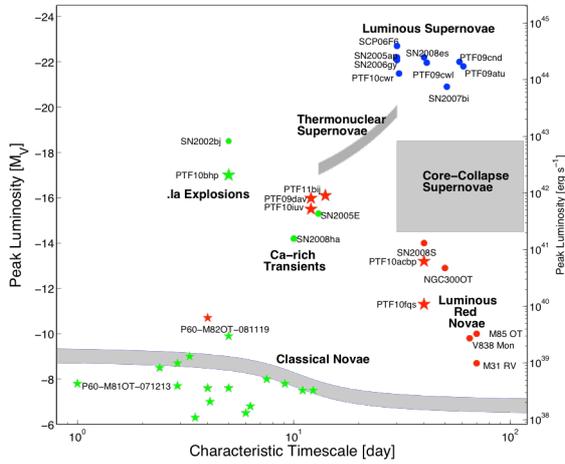

**Figure 1**. MMRD relationship with transients, from [3]

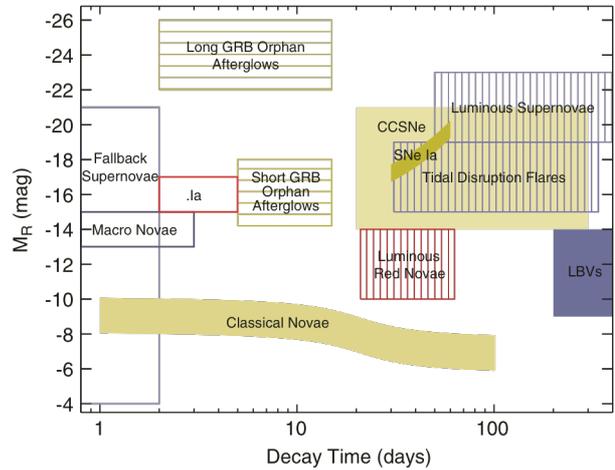

**Figure 2.** Luminosity vs. decay time plot for transients, from [4]

A similar situation confronted stellar population studies after studies of galaxies showed that three parameters were important for defining the stellar population of a galaxy: age, metal abundance, and star formation history. This led to the concept of the 3-dimensional population box [5]. The same situation appears to apply to transients, where the identification of a third parameter, in addition to the luminosity and decline time, would be useful in distinguishing between different classes of transients.

Spectroscopy of transients offers easily obtained information that allows kindred objects to be identified and differentiated among the different types of objects. Thus, the spectroscopic characteristics of a transient are arguably the best way to determine what the object is and how it may be similar to or different from other transients. This same problem has been addressed by the quasar community, who have applied principal component analysis (PCA) to quasar spectra [6,7] to define those properties most useful in discriminating between different quasar properties and types.

Our understanding of transients would be greatly improved by subjecting the spectra of all such objects, i.e., novae, supernovae, stellar mergers, LBVs, [Ca II] transients, etc., to PCA. The breakdown of the main spectral features into principal components, or eigenvectors, could then be used to identify the largest PC (pc1, or eigenvector1 to use the nomenclature of the quasar community) that could be used as the third parameter to create a 3-dimensional 'transient cube', as shown in figure 3. Pc1 might or might not correspond to some clear spectroscopic property such as the continuum slope or intensity of H-alpha or strength of narrow absorption lines, etc., however it almost certainly would better enable transients that are similar to be defined and differentiated from other types of transients. Classifying transients by plotting them on a diagram of luminosity vs. a principal component, such as pc1, would be much more instructive than the current practice of showing their position in a plot of luminosity vs. rate of brightness decline.

## 3. Novae Ejecta Geometry

Many novae, especially those having slower declines in brightness from maximum, show variation in visible brightness of several magnitudes over timescales of weeks. Multiple secondary maxima sometimes occur in addition to short-lived up-and-down fluctuations in brightness, i.e., jitter, that last a few days [8]. These are no doubt due to changing conditions in the ejecta, but no specific explanation for this behaviour has emerged.

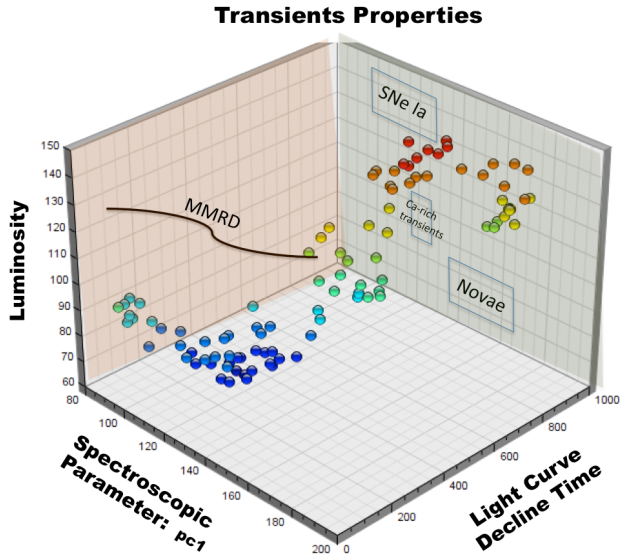

**Figure 3**. Consideration of a third parameter that characterizes the spectrum of a transient would provide important information that better allows identification of the class of the transient.

Many novae form dust that produces an infrared continuum, and high density regions in the ejecta protected from high UV radiation so that molecular species are plentiful are believed to define the locale of the dust formation. Repeated observations of postoutburst novae by the *Swift* satellite have detected X-ray emission in most novae [9]. Initially, the X-rays have hard 2-10 keV energies which progressively weaken until eventually soft X-rays emerge as the expanding ejecta reveal the hot post-TNR WD surface.

All three features relating to the brightness variations, dust formation, and the evolution of the X-ray emission may be explained by constraining various ejecta parameters which do evolve over time due to their expansion. However, no coherent explanation of all of these phenomena exists that relates them to each other and explains how they differ so much between different novae. We propose here a model, admittedly speculative until detailed calculations can confirm the model, in which fluctuations in brightness, the formation of dust, and observed X-ray emission from novae in postoutburst decline all follow naturally from ejecta that are formed from an ensemble of dense clouds of different sizes that are ballistically ejected by the TNR.

In our model the small clouds collide with each other, emitting hard X-rays as they rapidly dissipate to form a low density ambient medium. Less common large clouds have dense cores surrounded by a photosphere that provides protection for the outer dense regions where dust can form. When one of the large clouds emerges from the lower density ambient region of ejecta its radiation will produce a variation in brightness of the ensemble. The large variation in characteristics of different novae can then be understood in terms of a differing size distribution of ejected clouds from object to object. If we parameterize the size distribution of the clouds in terms of a power law distribution of slope β, the β-parameter dictates the fundamental characteristics of the ejecta.

The spectra of postoutburst novae show clear observational evidence for two very different components of gas. One component is responsible for the narrow low ionization, transient heavy element absorption (*thea*) systems observed in the weeks following outburst [10,11], and the other is responsible for the much broader absorption components that often include higher ionization species such as He I and II in the visible, and C IV, N V, etc., in the UV [12]. Although narrow and broad absorption line components frequently appear in the spectrum together they just as often have different times of appearance. Two parameters that demonstrate the separateness of the broad vs. narrow components of gas are the distinctly different radial velocities and appearance times of the two types of systems. An illustration of this is shown in figure 4 from excellent data obtained and reduced by Fred Walter with the high resolution, R=80,000, Chiron spectrograph [13] on the CTIO SMARTS 1.5m telescope, that is posted on Walter's website from his 2014 January AAS poster[1] and reproduced here with his permission.

Several different resolved, narrow line Na I D absorption components having radial velocities of $v_{rad}$ ~ -560 & -1,150 km/s are shown in figure 4, first detectable on JDs 6633 & 6639. Notably broad He I λ5876 absorption appears on JD 6645, having been present very weakly at lower radial velocities on several days the previous week. The broad He I line negative radial velocities of $v_{rad}$ < -1,500 km/s are higher than those of the resolved Na I D lines. More significantly, the radial velocity evolution of the He I absorption is rapidly variable from day to day, rather different from the slow progression of Na I D line radial velocities that increase steadily with time. A third Na I D *thea* system appears on JD 6649, quite broad for a *thea* system and having the same negative radial velocity $v_{rad}$ ≈ -1,530 km/s as He I. Appearing later than He I 5876 this Na I D absorption may originate in the same general component of gas as the He I absorption.

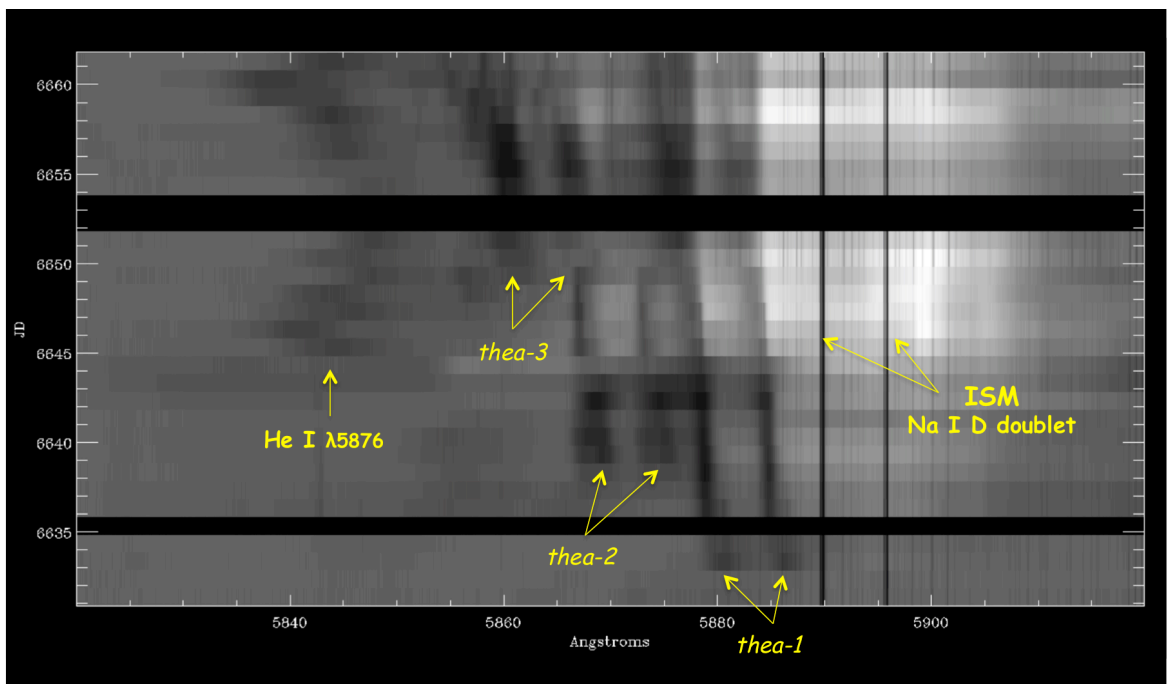

**Figure 4.** Spectral evolution of Nova V1369 Cen/2013 in the Na I D line region. Three resolved absorption components of Na I D appear at separate times, each of which progressively increase outwardly in radial velocity. A broader P Cygni absorption feature due to He I λ5876 appears separately, showing a different velocity behavior (courtesy of F. Walter)

Strong evidence exists for a component with very large density inhomogeneities that are ballistically ejected by the outburst. Specifically, (a) most *thea* systems exhibit O I λ7773 in absorption which, as a quintet transition, requires high densities $n>10^{12}$ cm$^{-3}$ for its excitation [14], and the radial velocities and lifetimes of the absorption systems require these densities to persist to distances of 1 au from the binary system, (b) narrow *thea* absorption systems [11] often maintain constant radial velocities for weeks, and (c) the constancy of the radial velocities of the narrow absorption lines during intervals of brightness fluctuations that approach several magnitudes amplitude, e.g., V5588 Sgr/2011 [15] and V1369 Cen/2013 [16], argues against formation of either the absorption or emission lines in a wind because substantial brightness variations should be accompanied by observable variations in the wind structure, yet these are not observed.

---

[1] http://www.astro.sunysb.edu/fwalter/SMARTS/NovaAtlas/v1369cen/V1369Cen.pptx.pdf

## 4. Ejecta as Ensemble of Clouds

We propose here that the characteristics of novae spectra and their variations in brightness may be explained by ejection of an ensemble of globules, or density inhomogeneities, created by the TNR. The internal structure of a freely expanding cloud does not yield a general analytical solution for density and temperature although there are self-similar solutions for specific conditions [17,18,19]. Numerical simulations have been carried out that validate the self-similar solutions [20], although the majority of these have been 2-dimensional simulations. In their analytical study of blast waves Ostriker & McKee [21] derive relationships for a variety of conditions, some of which are applicable to novae outbursts. In specific applications to supernova remnants Truelove and McKee [20] couple analytical results with numerical simulations that validate many of the analytical approximations used in describing blast waves, and we draw upon their results.

As has been demonstrated from the numerical simulations of the nova outburst [22] and of core collapse supernovae [23,24] various instabilities produce strong density inhomogeneities that have a range of sizes. The smaller globules quickly dissipate by expansion to merge together to form a diffuse medium that constitutes a lower density, more homogeneous, somewhat confining, component of the ejecta. The larger globules retain their structural integrity within the diffuse component, with their cores surviving intact for decades as their outer layers slowly dissipate. In particular, the Ellinger et al. [23] calculations confirm that larger globules will have lifetimes of decades even when they are not confined.

The rapidly increasing energy generation on the WD surface that drives the thermonuclear runaway results in ejection of surface gas that is ballistic and very inhomogeneous. The ensemble of knots are all assumed to be of initially homogeneous density $n_o \sim 10^{26}$ cm$^{-3}$ and temperature $T_o \sim 10^{7.5}$ K, somewhat less than the partially degenerate TNR conditions above the WD surface after degeneracy has been removed by the TNR [25,26]. We generalize this situation by assuming the inhomogeneities to be globules ejected with sizes between initial fiducial minimum and maximum radii $r_1$ and $r_2 \gg r_1$ and with a power law distribution such that the fraction $f(r_i)$ of globules ejected with initial radius between $r_i$ and $r_i + dr_i$ is given by

$$f(r_i) = (\beta - 1)/r_1 \times (r_i/r_1)^{-\beta} , \tag{1}$$

where $\beta > 1$, subject to the normalization

$$\int_{r_1}^{r_2} f(r_i) \, dr_i = 1 . \tag{2}$$

We assume the globules to be ejected from the WD with velocity $v_{ej}$ into a vacuum such that they immediately expand at the sound speed of their outer layers, $v_s \sim 700$ km/s, corresponding to their initial temperature $T_o \sim 10^{7.5}$ K. This is a worst case assumption for the lifetimes of the globules because in reality the smallest clumps will expand and coalesce to form a lower density ambient, confining medium. The globule expansion speed decreases steadily as adiabatic cooling lowers the temperature in the clouds. Adiabatic expansion of clouds leads to a decrease in density and temperature following the relation $T \propto n^{2/3}$ for specific heat ratio $\gamma = 5/3$.

Pressure confinement of the inner gas in blobs from the overlying gas produces a velocity, density, and temperature structure where the expansion velocity increases outward linearly in the 'linear velocity approximation' [21], resulting in inner plateaus of density and temperature [20]. The more rapidly expanding outer layers of each cloud produce an ambient lower density envelope that consists of two components: (a) hot, shocked gas created by the collision of the outer coronae of (primarily smaller) clouds with each other, and (b) a cold component of higher density adiabatically cooled gas in the intermediate regions of the knots that have not yet been subjected to the reverse shocks produced by the colliding clouds [21,27]. For our purposes we define the effective outer boundary of each globule, or inhomogeneity, at time $t_g$ to be the radius $r_g$ to which the initial radius $r_i$ has expanded at the steadily decreasing sound speed resulting from adiabatic cooling due to the expansion,

$$r_g(t) = r_i + \int_0^{t_g} v_s(t)\, dt, \qquad (3)$$

with the sound speed $v_s=(2kT_g/m_p)^{1/2}$.

The radial distribution of density for expanding globules has been published for few numerical simulations [28], however an analytic expression for the density distribution in the early "ejecta dominated (ED)" phase of free expansion have been derived for self-similar solutions by Truelove & McKee [20]. A moderate density gradient occurs for the inner region of the globule, reflecting the higher interior sound speed compared to the steadily decreasing expansion speed of the clump, that becomes steep close to the outer radius $r_g$. A mean density and temperature of a globule can be defined that are generally fall within factors of 3-4 of the central values of these parameters within each cloud. The time-dependent mean temperature and density for an expanding blob can then be found from equation (3) using the adiabatic relations and mass conservation, such that $T_g \propto n_g^{2/3} \propto r_g^{-2}$.

The time required for a globule of initial radius $r_i$ to expand to the point where its characteristic mean density has declined to $n_g$ is

$$t_g = \int_{r_i}^{r_g} v_s(r)^{-1}\, dr \ = r_i \times [(n_o/n_g)^{2/3} - 1]/[2v_s(T_o)]. \qquad (4)$$

We assume clouds to be ejected with the same initial density and temperature, $n_o=10^{26}$ cm$^{-3}$ and $T_o=10^{7.5}$ K, following the TNR on the WD surface. The resulting evolutionary tracks of clouds in density and radius for different initial radii $r_i$ are shown in figure 5. The curves are labeled with the time required for the globule to evolve to that value. For all cloud mean densities $n_g<10^{19}$ cm$^{-3}$ adiabatic expansion forces temperatures to be quite low, $T_g<10^3$ K, except for outermost layers heated by collisional shocks and any radiative heating of the globules.

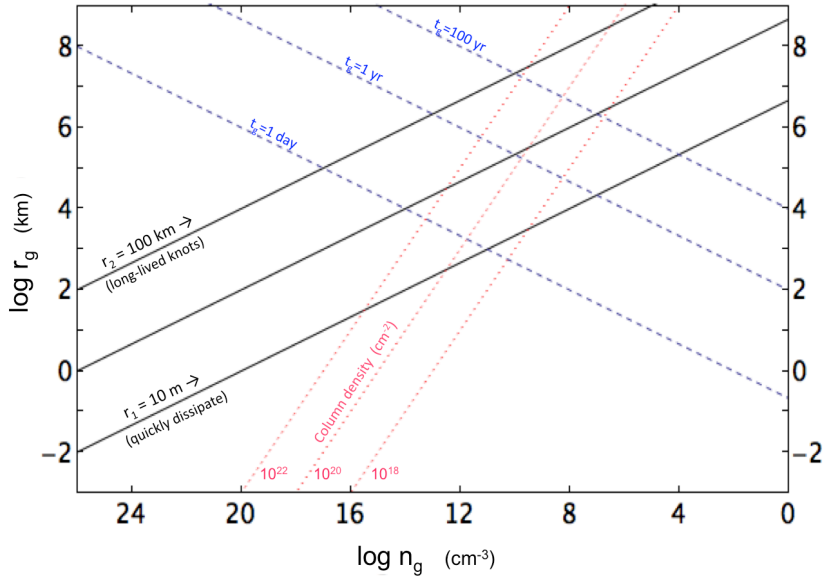

**Figure 5.** The evolution in size, density, and column density for adiabatically expanding globules of different initial size.

We confine attention to knots with initial sizes between the limits $r_1=10$ m and $r_2=100$ km, which corresponds to the maximum size of inhomogeneities from KH instabilities found in 3-dimensional numerical hydrodynamic simulations of the outburst [22]. Figure 5 shows that the larger globules retain densities of ~$10^{10}$ cm$^{-3}$ for more than 100 yr while growing in size to ~$10^{12}$ cm and maintaining a column density of $10^{22}$ cm$^{-2}$. Such long-lived knots retaining high core densities are the prime candidates for the UV, optical, IR emission sources observed in spatially resolved novae shells. At the other extreme, the smaller clouds dissipate within one day to densities of ~$10^{11}$ cm$^{-3}$ with column densities of $10^{19}$ cm$^{-2}$ after expanding to sizes of $10^{8}$ cm. The latter size is comparable to the WD radius so the large population of small clouds merge with each other quickly to form an extensive circumbinary envelope, or interglobule region (IGR), of gas that, depending on the mass ejected, we speculate may become optically thick in the optical and UV, shielding the WD and innermost ejecta from direct observation.

The initial cloud size distribution function plays a key role in determining the spectral evolution of novae. *The basic characteristics of the ejecta will be dictated by the β-parameter*. A predominance of larger clouds will result in more narrow *thea* systems whereas a predominance of small clouds will favor the formation of broad P Cygni absorption that will transition to the forbidden emission line phase of the postoutburst spectrum.

There are observations that provide constraints on the value of β in equation (1). Ejected blobs that produce the saturated *thea* absorption lines must have a column density in Na I of at least ~$10^{15}$ cm$^{-2}$, which translates to a total gas column density of $<n_g r_g> \sim 10^{18}$ cm$^{-2}$ using recently compiled ejecta abundances [29]. The observed central absorption line residual intensities, i.e., depths, of *thea* systems indicate typically $r_g \sim \frac{1}{2} R_{phot}$, therefore such clouds should have core sizes that are neither significantly larger nor smaller than typical postoutburst effective photospheric radii of ~100 $R_\odot$ [30,31]. The lifetimes of such clouds should be of order 1-5 weeks, as is observed for most *thea* systems, and they should have temperatures that are consistent with observed velocity dispersions as low as ~35 km/s. Furthermore, the size distribution of ejected clouds should lead to no more than ~1-4 of them being detectable along the line of sight at any one time, as observed at any given time in the original FEROS novae sample [11].

The observation that novae have relatively few detectable *thea* systems constrains the size distribution of ejected clouds, particularly for the largest sizes. A flat size distribution produces more large, long-lived blobs whereas a steep distribution favors the lower density, more homogeneous shell component, resulting in its higher mass and optical depth so that it obscures the embedded large globules. The ejected globule size distribution as determined by the parameter β therefore controls the basic nature of the spectrum from having primarily P Cygni profiles prominent (β>3, homogeneous medium) to having prominent *thea* absorption features (β<2, primarily large discrete globules).

## 5. Brightness Fluctuations

Ejected globules can account for jitter re-brightening and fading during the optically thick IGR phase. The sudden appearance of a large globule or two that emerge from beneath the IGR effective photosphere will produce brightening and then fading from its expansion. Such brightness fluctuations may be correlated with the presence of individual narrow line *thea* absorption systems, and in fact just such a correlation has recently been reported to the author by F. Walter (2015, private communication). It does appear that short-term brightness fluctuations in postoutburst decline may appear preferentially in novae with multiple *thea* systems that have broader, more complex profiles.

## 6. Formation of Dust and X-ray Emission

Formation of dust in novae is likely to occur when the gas temperature falls below the condensation temperature of silicates and carbon, T~1,000 K, the density is sufficiently high for molecular formation and clustering to occur, n> $10^{10}$ cm$^{-3}$, and the region is free of UV radiation that would photodissociate molecules [32]. In the globule model advocated here hard X-ray emission (kT≈2-10 keV) is produced both within the high density, T~$10^{7}$ K cores of the globs and also more weakly in the lower density shocked IGR interfaces of colliding knots.

Relatively few novae have formed dust in the era of the SWIFT satellite, but one of them, V5668 Sgr/2015, formed dust four months after outburst during which time the SWIFT satellite detected a weak but increasing flux of both hard and soft X-rays during the dust formation interval [33]. Thus,

V5668 Sgr presents the interesting dichotomy of a nova showing simultaneously both dust and hard X-ray emission. Expanding globules do present favorable conditions for both phenomena. Favorable conditions for dust formation occur outside the globule core where adiabatic expansion cools the gas in regions where the density remains above $10^{12}$ cm$^{-3}$. It does require that the gas interior be sufficiently optically thick to the hot central X-ray+UV emitting core that it shields the outer layers so molecules can form. Ionizing radiation will not escape the globule core in this case, being degraded to lower energies. Separately, the outermost layers of knots, i.e., the low density shocked interface IGR region, will have sufficiently high temperatures to emit hard X-rays, although much weaker than in the much denser large globule cores. Thus, dust formation in ejected knots can be expected to occur in the intermediate density layers of globs that are sandwiched between the obscured X-ray emitting centers and outermost shocked layers. Protection from the central core ionizing and dissociating radiation needs to be provided the dust formation layers by the opacity of the inner gas. If sufficiently weak, the IGR X-ray emission should not present a problem for molecule formation in the dust layers. The dust layers cannot be shielded from IGR radiation by overlying optically thick gas because that gas would then constitute an effective photosphere for the blob, preventing the dust layers from being detected. A simplified schematic of this geometry is shown in figure 6, where a large globule and an optically thick IGR can provide conditions for dust formation in the presence of weakly emitting X-rays from regions of the shocked IGR.

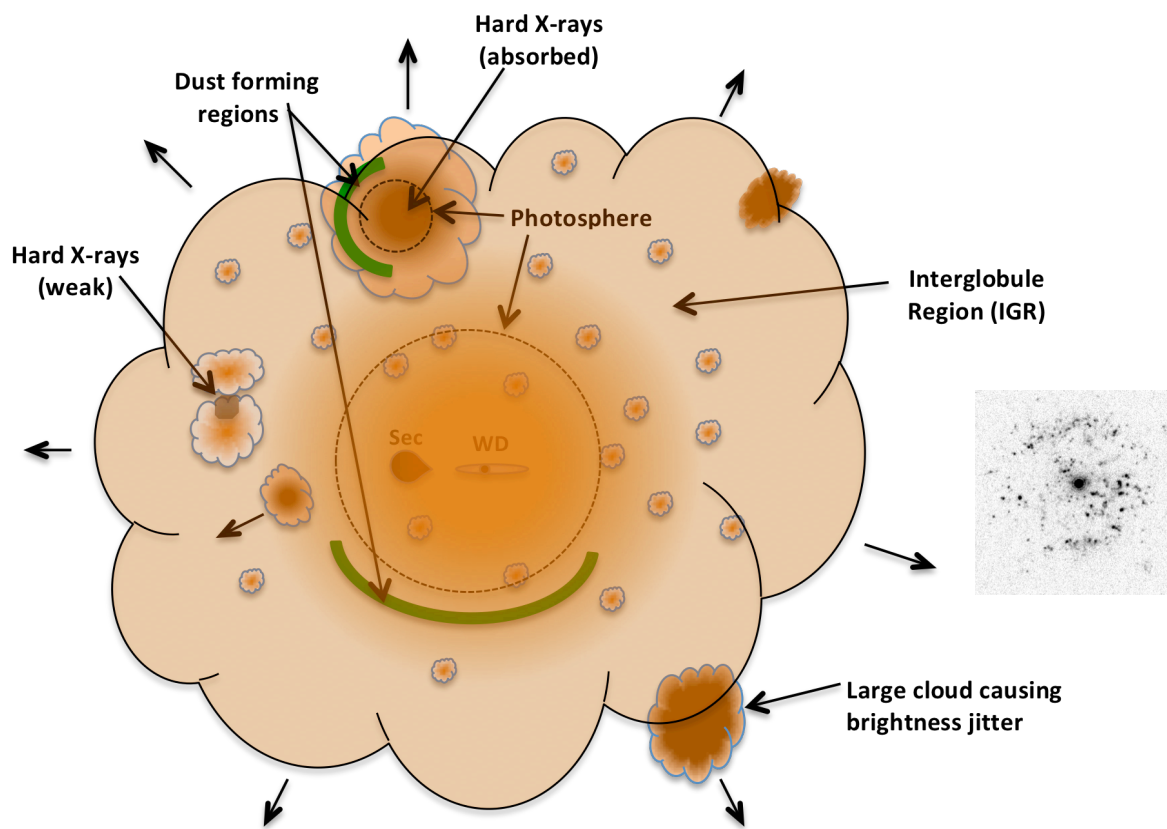

**Figure 6.** Idealized geometry of ejecta where dust and hard X-rays form in separate regions. The dust forms both in cold, moderately high density regions of large globules, protected from the ionizing radiation of the interior, and in the outer region of the larger IGR. The hard X-rays originate in the shocked IGR regions between colliding globules. Large clouds that are revealed as the IGR expands and dissipates produce brightness variations. The inset shows the resolved ejecta of the recurrent nova T Pyxidis [34].

The above situation leads to a number of different possibilities for explaining the formation of dust in novae and the presence or absence of X-rays when dust has formed. As can be seen from figure 5 the presence of gas with n> $10^{12}$ cm$^{-3}$ weeks or months after outburst requires knots with initial size greater than ~10 km. The total column density of such a knot will exceed $10^{21}$ cm$^{-2}$, so shielding from core X-ray emission is reasonable. A predominance of initially small globules will discriminate against copious high density regions in the ejecta, preventing significant dust formation. For those novae that do form dust the usual delay of weeks to months for dust formation may be due to the required build up of the cooler protective region of optically thick gas between the X-ray + UV emitting central core and the colder gas at larger radii where molecule and dust formation can take place. The size distribution function of ejected globules will therefore play a key role in the dust and X-ray evolution of novae in decline.

An indicator of the location of dust could come from the O I $\lambda$7773 quintet absorption profile whose formation, like that of dust, is confined to relatively high density regions with n>$10^{12}$ cm$^{-3}$. If when dust is present in novae ejecta the O I feature shows a broad P Cygni profile it would suggest that the dust has likely formed in the IGR region. Alternatively, if the O I $\lambda$7773 absorption is very narrow and part of a *thea* absorption system it would indicate that the dust is associated with a large globule(s).

**7. Summary**

Inhomogeneities are an important component of novae ejecta. They are observed to persist for many decades and the spectroscopic evidence for large density and ionization variations is strong. A fundamental question is whether the requisite orders of magnitude density fluctuations are created by hydrodynamic instabilities, e.g., Rayleigh-Taylor, from a 2-component flow of gas from the binary system or are created ab initio by the outburst from the ejection of extremely high density blobs of gas from the WD surface that promptly expand.

The observed relative strengths of both the emission and absorption components of O I $\lambda$7773/O I $\lambda$8446 require high densities in the emitting and absorbing gas for months following outburst, and they must have large size, initially exceeding 100 km if they are to be long-lived. Whatever creates the blobs, their size distribution dictates the nature of the spectral evolution of the ejecta, including X-ray and optical characteristics, and molecular and dust formation. Quantitative treatments of the adiabatic expansion of initially partially degenerate gas globules in conditions applicable to novae will be helpful. Frequent high resolution spectroscopic and photometric follow up of postoutburst novae over a range of wavelength regions can provide the data with which to compare the theoretical calculations to ascertain the geometry and physical conditions in the ejecta.

The author is grateful to Dr. Fred Walter for making his reduced data available. This research has been funded in part by grant HST-GO-13388.011-A.

**Acknowledgment**